\documentclass[aps,floats]{revtex4}
\usepackage{amsmath,amssymb}
\usepackage{graphicx,epsfig}

\begin{document}
\bibliographystyle {plain}

\def\oppropto{\mathop{\propto}} 
\def\opsimeq{\mathop{\simeq}}
\def\opoverderline{\mathop{\overline}}
\def\operarrow{\mathop{\longrightarrow}}
\def\opsim{\mathop{\sim}}

\def\fig#1#2{\includegraphics[height=#1]{#2}}
\def\figx#1#2{\includegraphics[width=#1]{#2}}


\title{ Eigenvalue method to compute the largest relaxation time of disordered systems } 


 \author{ C\'ecile Monthus and Thomas Garel }
  \affiliation{ Institut de Physique Th\'{e}orique, CNRS and CEA Saclay,
 91191 Gif-sur-Yvette, France}

\begin{abstract}
We consider the dynamics of finite-size disordered systems as defined by a master equation satisfying detailed balance. The master equation can be mapped onto a Schr\"odinger equation in configuration space, where the quantum Hamiltonian $H$ has the generic form of an Anderson localization tight-binding model. The largest relaxation time $t_{eq}$ governing the convergence towards Boltzmann equilibrium is determined by the lowest non-vanishing eigenvalue $E_1=1/t_{eq}$ of $H$ (the lowest eigenvalue being $E_0=0$). So the relaxation time $t_{eq}$ can be computed {\it without simulating the dynamics } by any eigenvalue method able to compute the first excited energy $E_1$. Here we use the 'conjugate gradient' method to determine $E_1$ in each disordered sample and present numerical results on the statistics of the relaxation time $t_{eq}$ over the disordered samples of a given size for two models : (i) for the random walk in a self-affine potential of Hurst exponent $H$ on a two-dimensional square of size $L \times L$, we find the activated scaling $\ln t_{eq}(L) \sim L^{\psi}$ with $\psi=H$ as expected; (ii) for the dynamics of the Sherrington-Kirkpatrick spin-glass model of $N$ spins, we find the growth $\ln t_{eq}(N) \sim N^{\psi}$ with $\psi=1/3$ in agreement with most previous Monte-Carlo measures. In addition, we find that the rescaled distribution of $(\ln t_{eq})$ decays as $e^{- u^{\eta}}$ for large $u$ with a tail exponent of order $\eta \simeq 1.36$. We give a rare-event interpretation of this value, that points towards a sample-to-sample fluctuation exponent of order $\psi_{width} \simeq 0.26$ for the barrier.

\end{abstract}

\maketitle

\section{ Introduction }

The non-equilibrium dynamics of disordered systems has been much
studied both experimentally and theoretically
(see for instance the reviews \cite{bouchaud,berthierhouches} 
and references therein).
In numerical simulations, the main limitation
is that the equilibrium time $t_{eq}(L)$ needed to converge
towards equilibrium for a finite system of linear size $L$
grows very rapidly with $L$.
Within the droplet scaling theory
proposed both for spin-glasses \cite{heidelberg,Fis_Hus} 
and for the directed polymer in a
random medium \cite{Fis_Hus_DP}, the non-equilibrium dynamics
 is activated with barriers scaling as a power law $B(L) \sim L^{\psi}$
with some barrier exponent $\psi>0$ that is independent
 of temperature and disorder strength.
The equilibrium time $t_{eq}(L)$ then grows as
\begin{eqnarray}
\ln t_{eq}(L) = B(L) \sim L^{\psi}
\label{defpsi}
\end{eqnarray}
This logarithmic scaling has been used to fit numerical data
for disordered ferromagnets \cite{Hus_Hen,puri,bray_hum} 
and spin-glasses \cite{huseSG,berthier}.
Other authors, both for disordered ferromagnets \cite{rieger_coarsening,henkel}
and spin-glasses \cite{kis,kat_cam} 
prefer a scenario corresponding to logarithmic barriers 
$B(L) \sim z(T,\epsilon) \ln L$, so that the equilibrium time $t_{eq}(L)$ scales as
a power-law
\begin{eqnarray}
 t_{eq}(L) = e^{B(L)} \sim L^{z(T,\epsilon)}
\label{lz}
\end{eqnarray}
where the exponent $z(T,\epsilon)$ is non-universal and 
depends on the temperature $T$ as well as on the disorder strength $\epsilon$.
In the field of directed polymers or elastic lines in random media,
the fit based the algebraic form of Eq. \ref{lz} used
 initially by many authors \cite{DP_alge}
has been now excluded by more recent work \cite{rosso,leticia,noh}, and has been interpreted
as an artefact of an initial transient regime \cite{leticia,noh}.
The reason why the debate between the two possibilities of Eqs \ref{defpsi} and \ref{lz}
 has remained controversial over the years for many interesting disordered models is 
that the equilibrium time $t_{eq}(L)$ grows numerically so rapidly with $L$
that $t_{eq}(L)$ can be reached at the end of dynamical simulations only for rather small 
system sizes $L \leq L_{max}$.
For instance, in Monte-Carlo simulations of 2D or 3D random ferromagnets
 \cite{Hus_Hen,puri,bray_hum,rieger_coarsening,henkel,sicilia} or spin-glasses 
\cite{huseSG,berthier,kis,kat_cam},
the maximal equilibrated size is usually only of order $L_{max} \sim 10$ lattice spacings.
Even faster-than-the-clock Monte Carlo 
algorithms  \cite{algoBKL}, where each iteration leads to a movement, 
 become inefficient 
 because they face the 'futility' problem \cite{werner} :
the number of different configurations visited during the simulation 
remains very small with respect to the accepted moves, i.e.
 the system visits over and over again 
the same configurations within a given valley before it is able to
escape towards another valley.
A recent proposal to improve significantly Monte Carlo simulations
 of disordered systems
consists in introducing some renormalization ideas \cite{chanal}.

Taking into account these difficulties, a natural question is whether it could
be possible to obtain
informations on the equilibrium time $t_{eq}(L)$
{ \it without simulating the dynamics}.
In previous works \cite{rgmaster,rgu2d}, we have proposed for instance
to study the flow of some strong disorder renormalization procedure acting
on the transitions rates of the master equation.
However this approach is expected to become asymptotically exact only
if the probability distribution of renormalized transitions rates
flows towards an 'infinite disorder' fixed point, i.e. only
 for the activated scaling of Eq. \ref{defpsi}.
In the present paper, we test another strategy to compute $t_{eq}$
which is a priori valid for any dynamics defined by a
master equation satisfying detailed balance : it is based on the computation 
of the first excited energy $E_1$ of the quantum Hamiltonian $H$
 that can be associated to
the master equation. This approach makes no assumption on the nature
of the dynamics and is thus valid both for activated or non-activated dynamics
(Eqs \ref{defpsi} or \ref{lz}). 
The mapping between continuous-time stochastic dynamics with detailed balance
and quantum Schr\"odinger equations is of course very well-known and can be found
in most textbooks on stochastic processes (see for instance 
\cite{gardiner,vankampen,risken}). However, since it is very often explained
on special cases, either only in one-dimension, or only for continuous space,
 or only for Fokker-Planck equations, we stress here that this mapping is valid for any master equation satisfying detailed balance 
(see more details in section \ref{secquantum}).
In the field of disordered systems, this mapping has been very much used for
one-dimensional models (see the review \cite{jpbreview} and references
therein, as well as more recent works \cite{laloux,golosovloc,texier}),
but to the best of our knowledge, it has not been used in higher dimension,
 nor for many-body problems.
In the field of many-body dynamics without disorder,
 this mapping has been already used as a numerical tool to measure very precisely 
the dynamical exponent $z$ of the two dimensional 
 Ising model at criticality \cite{Night_Blo}.

The paper is organized as follows.
In section \ref{secquantum}, we recall how the master equation can be mapped
onto a Schr\"odinger equation in configuration space, 
and describe how the equilibrium time $t_{eq}$ can be obtained
 from the associated quantum Hamiltonian.
We then apply this method to two types of disordered models : 
section \ref{secRW} concerns 
 the problem of a random walk in a two-dimensional self-affine potential, and  section \ref{secSK} is devoted to the the dynamics of the Sherrington-Kirkpatrick spin-glass model. Our conclusions are summarized in section 
\ref{secconclusion}.

\section{ Quantum Hamiltonian associated to the Master Equation }

\label{secquantum}

\subsection{ Master Equation satisfying detailed balance} 

In statistical physics, it is convenient to consider
continuous-time stochastic dynamics defined by a master equation of the form
\begin{eqnarray}
\frac{ dP_t \left({\cal C} \right) }{dt}
= \sum_{\cal C '} P_t \left({\cal C}' \right) 
W \left({\cal C}' \to  {\cal C}  \right) 
 -  P_t \left({\cal C} \right) W_{out} \left( {\cal C} \right)
\label{master}
\end{eqnarray}
that describes the  the evolution of the
probability $P_t ({\cal C} ) $ to be in  configuration ${\cal C}$
 at time t.
The notation $ W \left({\cal C}' \to  {\cal C}  \right) $ 
represents the transition rate per unit time from configuration 
${\cal C}'$ to ${\cal C}$, and 
\begin{eqnarray}
W_{out} \left( {\cal C} \right)  \equiv
 \sum_{ {\cal C} '} W \left({\cal C} \to  {\cal C}' \right) 
\label{wcout}
\end{eqnarray}
represents the total exit rate out of configuration ${\cal C}$.
Let us call $U(C)$ the energy of configuration ${\cal C}$.
To ensure the convergence towards Boltzmann equilibrium at temperature $T$
in any finite system
\begin{eqnarray}
P_{eq}({\cal C}) = \frac{ e^{- \frac{U({\cal C})}{T}} }{Z}
\end{eqnarray}
where $Z$ is the partition function
\begin{eqnarray}
Z = \sum_{\cal C}  e^{- \frac{U({\cal C})}{T}} 
\label{partition}
\end{eqnarray}
it is sufficient to impose the detailed-balance property
\begin{eqnarray}
 e^{-  \frac{U({\cal C })}{T} }  W \left( \cal C \to \cal C '  \right)
= e^{-  \frac{U({\cal C' })}{T} }  W \left( \cal C' \to \cal C   \right)
\label{detailed}
\end{eqnarray}

\subsection{ Mapping onto a Schr\"odinger equation in configuration space}

As is well known (see for instance \cite{gardiner,vankampen,risken})
 the master equation operator can be transformed into a symmetric operator
 via the change of variable
\begin{eqnarray}
P_t ( {\cal C} ) \equiv 
   e^{-  \frac{U(\cal C )}{2T} } \psi_t  ( {\cal C} )
\label{relationPpsi}
\end{eqnarray}
The function $\psi_t  ( {\cal C} )$ then satisfies an imaginary-time  Schr\"odinger
equation
\begin{eqnarray}
\frac{ d\psi_t \left({\cal C} \right) }{dt} = -H \psi_t \left({\cal C} \right)
\label{Hquantum}
\end{eqnarray}
where the quantum Hamiltonian has the generic form of an Anderson localization
 model in configuration space
\begin{eqnarray}
 H = \sum_{\cal C } \epsilon \left( {\cal C} \right) \vert {\cal C} > < {\cal C } \vert
+ \sum_{{\cal C},{\cal C '}}  V({\cal C} , {\cal C' })
 \vert {\cal C} > < {\cal C' } \vert
\label{tight}
\end{eqnarray}
The on-site energies read
\begin{eqnarray}
 \epsilon \left( {\cal C} \right) = W_{out} \left( {\cal C} \right)
\label{eps}
\end{eqnarray}
whereas the hopping terms read
\begin{eqnarray}
 V({\cal C} , {\cal C' })=- e^{-  \frac{(U(\cal C' )-U(\cal C ))}{2T} } W \left( \cal C' \to \cal C   \right)
\label{hopping}
\end{eqnarray}

\subsection{ Specific choices for the detailed balance dynamics}

To have the detailed balance of Eq. \ref{detailed}, it is convenient to 
rewrite the rates in the following form
\begin{eqnarray}
W \left( \cal C \to \cal C '  \right)
= \delta_{<\cal C, \cal C' >} 
\   e^{-  \frac{(U({\cal C' })-U({\cal C }))}{2T} } e^{- S (\cal C , \cal C ' )}
\label{Wratesgeneral}
\end{eqnarray}
where $\delta_{<\cal C, \cal C' >} $ means that the two configurations are
related by an elementary dynamical move, and where $S (\cal C , \cal C ' )$ is an arbitrary symmetric function : 
$S (\cal C , \cal C ' ) =S (\cal C' , \cal C  )  $.

\subsubsection{ Simplest choice $S ({\cal C , \cal C '} )=0$ }

To have the detailed balance property of Eq. \ref{detailed}, the simplest 
choice in Eq. \ref{Wratesgeneral} corresponds to $S ({\cal C , \cal C '} )=0$ 
\begin{eqnarray}
W \left( \cal C \to \cal C '  \right)
= \delta_{<\cal C, \cal C' >} 
\   e^{-  \frac{(U({\cal C' })-U({\cal C }))}{2T} } 
\label{WratesU}
\end{eqnarray}
Then the hopping terms of the quantum Hamiltonian are simply
\begin{eqnarray}
 V({\cal C} , {\cal C' })=-  \delta_{<\cal C, \cal C' >}
\label{hoppingsimple}
\end{eqnarray}
i.e. the non-vanishing hopping terms are not random, but take the same constant value $(-1)$ as in usual Anderson localization tight binding models. 
 The on-site energies are random and read 
\begin{eqnarray}
 \epsilon \left( {\cal C} \right) = \sum_{\cal C'} \delta_{<\cal C, \cal C' >} 
\   e^{-  \frac{(U(\cal C' )-U(\cal C ))}{2T} } 
\label{epssimple}
\end{eqnarray}

\subsubsection{ Metropolis choice  }
 
In numerical simulations, one of the most frequent choice corresponds to the
 Metropolis transition rates
\begin{eqnarray}
W \left( \cal C \to \cal C '  \right)
= \delta_{<\cal C, \cal C' >} 
{\rm min} \left[1, e^{-  \frac{(U({\cal C' })-U({\cal C }))}{T}} \right]
\label{metropolis}
\end{eqnarray}
In Eq. \ref{Wratesgeneral}, this corresponds to the choice
\begin{eqnarray}
S({\cal C , \cal C '} ) =    \frac{\vert U({\cal C' })-U({\cal C })\vert }{2 T} 
\label{Smetropolis}
\end{eqnarray}
In the quantum Hamiltonian, the hopping terms then read
\begin{eqnarray}
 V^{metropolis}({\cal C} , {\cal C' })=- \delta_{<\cal C, \cal C' >}
e^{-  \frac{\vert U({\cal C' })-U({\cal C })\vert }{2 T}}
\label{hoppingmetro}
\end{eqnarray}
and the on-site energies are given by
\begin{eqnarray}
 \epsilon^{metropolis} \left( {\cal C} \right) = \sum_{\cal C'} \delta_{<\cal C, \cal C' >} 
\   {\rm min} \left[1, e^{-  \frac{(U({\cal C' })-U({\cal C }))}{T}} \right]
\label{epsmetro}
\end{eqnarray}

\subsection{ Properties of the spectrum of the quantum Hamiltonian $H$ }

Let us note $E_n$ the eigenvalues of $H$ and $\vert \psi_n>$ the associated normalized eigenvectors
\begin{eqnarray}
H  \vert  \psi_n > && = E_n \vert \psi_n> \\
\sum_{\cal C} \vert \psi_n({\cal C}) \vert ^2 && =1
\label{spectreH}
\end{eqnarray}
The decomposition onto these eigenstates of the evolution operator $e^{-t H}$
\begin{eqnarray}
 <{\cal C} \vert e^{-t H}  \vert {\cal C}_0> = 
 \sum_n e^{- E_n t} \psi_n({\cal C})\psi_n^*({\cal C}_0)
\label{spectreHexp}
\end{eqnarray}
yields the following expansion for the conditional probability 
$P_t \left( {\cal C} \vert {\cal C}_0\right)$ to be in configuration ${\cal C}$ at $t$
if one starts from the configuration ${\cal C}_0$ at time $t=0$
\begin{eqnarray}
P_t \left( {\cal C} \vert {\cal C}_0\right) =
   e^{-  \frac{U({\cal C} )-U({\cal C}_0 )}{2T} } <{\cal C} \vert e^{-t H}  \vert {\cal C}_0>
= e^{-  \frac{U({\cal C} )-U({\cal C}_0 )}{2T} }
\sum_n e^{- E_n t} \psi_n({\cal C})\psi_n^*({\cal C}_0)
\label{expansionP}
\end{eqnarray}

The quantum Hamiltonian $H$ has special properties
that come from its relation to the dynamical master equation :

(i) the ground state energy is $E_0=0$, and the corresponding
eigenvector is given by

\begin{eqnarray}
\psi_0 ({\cal C}) = \frac{ e^{- \frac{U({\cal C})}{2T}} }{\sqrt Z}
\label{psi0}
\end{eqnarray}
where $Z$ is the partition function of Eq. \ref{partition}.

This corresponds to the convergence towards the Boltzmann equilibrium in Eq. \ref
{relationPpsi} for any initial condition ${\cal C}_0$
\begin{eqnarray}
P_t \left( {\cal C} \vert {\cal C}_0\right)
\opsimeq_{t \to + \infty}  e^{-  \frac{U({\cal C} )-U({\cal C}_0 )}{2T} }
\psi_0({\cal C})\psi_0^*({\cal C}_0) = \frac{e^{- \frac{U({\cal C})}{T}}}{Z} = P_{eq}({\cal C})
\label{CVeqP}
\end{eqnarray}

(ii) the other energies $E_n>0$ determine the relaxation towards equilibrium.
In particular, the lowest non-vanishing energy $E_1$
determines the largest relaxation time $(1/E_1)$ of the system 
\begin{eqnarray}
P_t \left( {\cal C} \vert {\cal C}_0\right) - P_{eq}({\cal C})
\opsimeq_{t \to + \infty} e^{- E_1 t}  e^{-  \frac{U({\cal C} )-U({\cal C}_0 )}{2T} }
\psi_1({\cal C})\psi_1^*({\cal C}_0) 
\label{CVeqP1}
\end{eqnarray}
Since this largest relaxation time represents the 'equilibrium time', 
i.e. the characteristic time needed to converge towards equilibrium, we will use the following
notation from now on
\begin{eqnarray}
t_{eq} \equiv \frac{1}{E_1}
\label{deftaueq}
\end{eqnarray}

The conclusion of this section is thus that
 the relaxation time $t_{eq}$ can be computed without simulating the dynamics
by any eigenvalue method able
to compute the first excited energy $E_1$ of the quantum Hamiltonian $H$
(where the ground state is given by Eq. \ref{psi0} and has for eigenvalue $E_0=0$). 
In the following subsection, we describe one of such methods called the 
 'conjugate gradient' method.

\subsection{ Conjugate gradient method in each sample to compute $E_1$ }

\label{secconjugate}

The 'conjugate gradient method' has been introduced as an iterative algorithm 
to find the minimum of functions of several variables with much better convergence
properties than the 'steepest descent' method \cite{shewchuk,golub}.
It can be applied to find the ground state eigenvalue and the associated eigenvector
by minimizing the corresponding Rayleigh quotient \cite{bradbury,nightingaleCG}
\begin{eqnarray}
R \equiv \frac{<v \vert H \vert v>}{<v \vert  v>}
\label{rayleigh}
\end{eqnarray}
The relation with the Lanczos method to solve large sparse eigenproblems
is discussed in the chapters 9 and 10 of the book \cite{golub}.
In the following, we slightly adapt the method described in 
\cite{bradbury,nightingaleCG} concerning the ground state $E_0$
to compute instead the first excited energy $E_1$ : the only change is that
the  Rayleigh quotient has to be minimized within the space orthogonal to the ground state.

In the remaining of this paper, we apply this method to various disordered models
to obtain the probability distribution of the equilibrium time $t_{eq}(L)$ over the samples
of a given size $L$.
  More precisely, since the appropriate variable 
is actually the equilibrium barrier defined as
\begin{eqnarray}
\Gamma_{eq} \equiv \ln t_{eq} = - \ln E_1
\label{defGammaeq}
\end{eqnarray}
we will present numerical results for the probability distribution $Q_L(\Gamma_{eq})$
for various sizes $L$.

 \section{ Random walk in a two-dimensional self-affine potential }

\label{secRW}

\begin{figure}[htbp]
 \includegraphics[height=6cm]{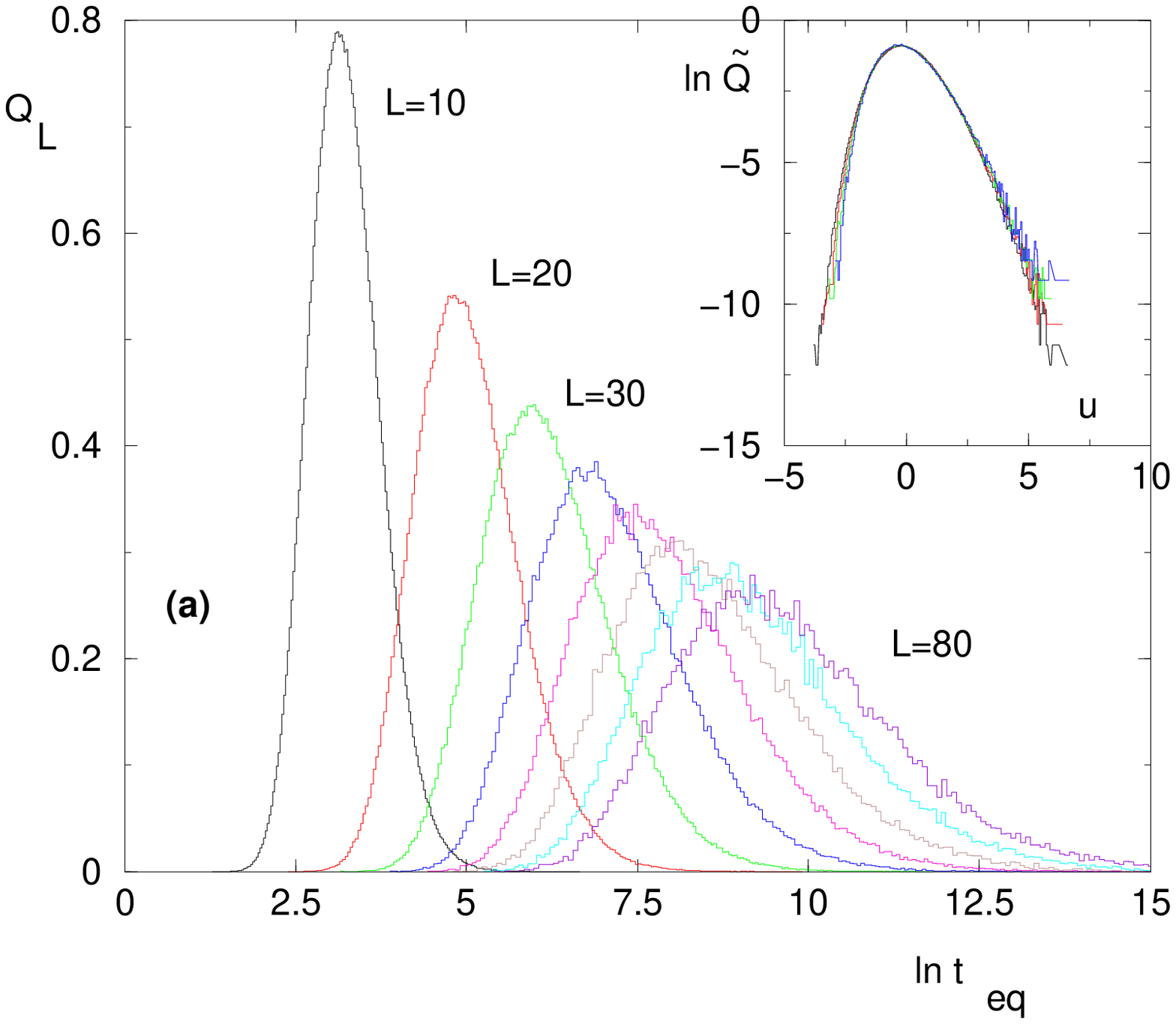}
\hspace{2cm}
 \includegraphics[height=6cm]{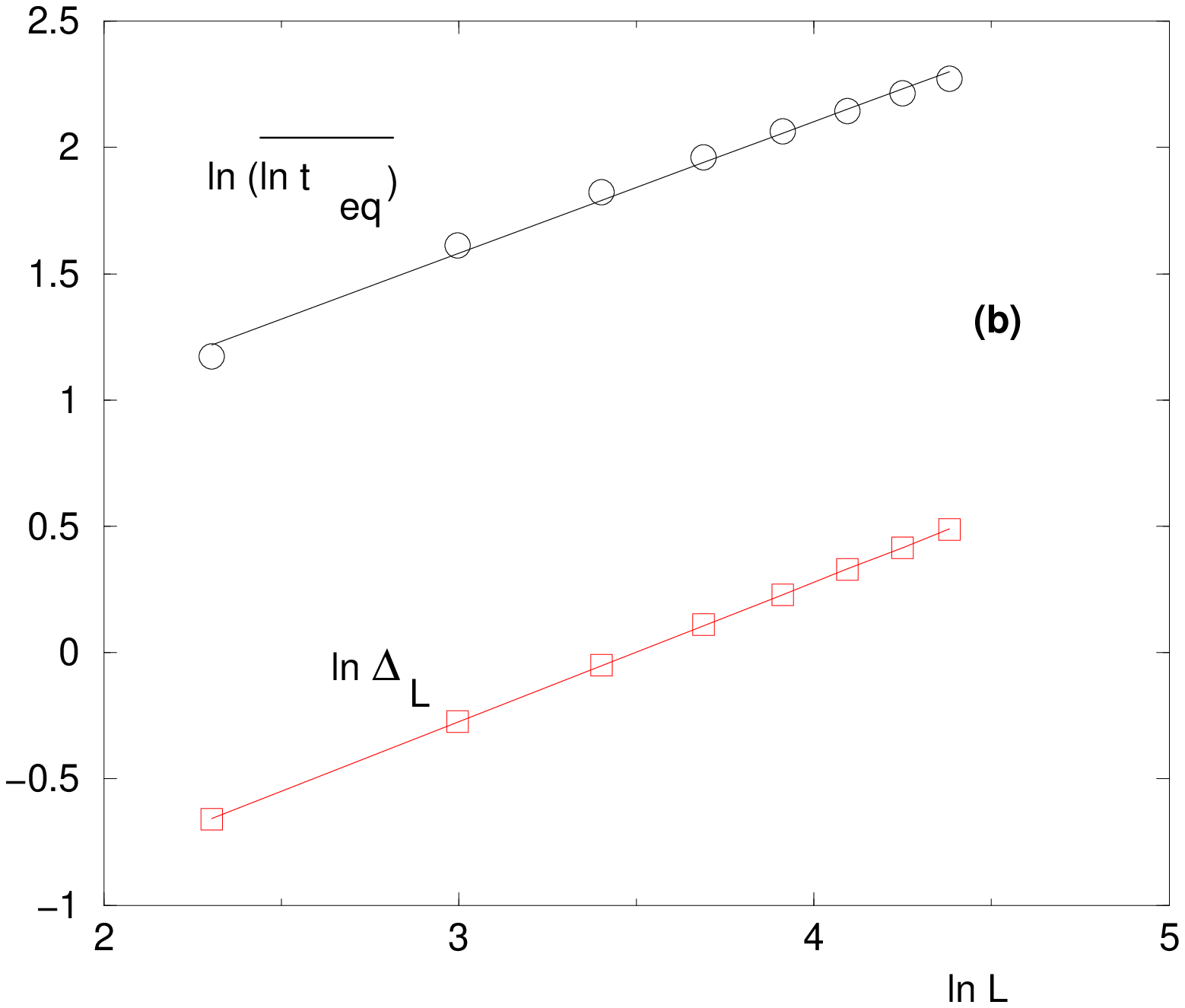}
\vspace{1cm}
\caption{ (Color on line) 
 Statistics of the equilibrium time $t_{eq}$ over the disordered samples of sizes $L^2$
 for the random walk in a two-dimensional self-affine random potential of Hurst exponent
 $H=0.5$ :
(a) Probability distribution 
 $Q_{L}(\Gamma_{eq}=\ln t_{eq})$ for $L=10,20,30,40,50,60,70,80$
(Inset : the corresponding distributions $\tilde Q (u)$
of the rescaled variable 
$u \equiv (\Gamma_{eq}-\overline{\Gamma_{eq}}(L) )/\Delta(L) $ 
are shown in log scale for $L=10,20,30,40$);
(b) the log-log plots of the disorder-average $ \overline{\Gamma_{eq}}(L)=
  \overline{\ln t_{eq} }(L) $ and of the width $\Delta(L)$
corresponds to the barrier exponent $\psi=H=0.5$ (Eq. \ref{psiboth}) }
\label{figu2d}
\end{figure}

In this section, we apply the method of the previous section
to the continuous-time random walk of a particle
 in a two-dimensional self-affine quenched random potential of Hurst exponent $H=0.5$.
Since we have studied recently in \cite{rgu2d} the very same model via some strong disorder
renormalization procedure, we refer the reader to \cite{rgu2d} and references therein
for a detailed presentation of the model and of the numerical method to generate the random potential. Here we simply recall what is necessary for the present approach.

We consider finite two-dimensional lattices of sizes $L\times L$.
 The continuous-time random walk in the random potential $U(\vec r)$
is defined by the master equation
\begin{eqnarray}
\frac{ dP_t \left({\vec r} \right) }{dt}
= \sum_{\vec r \ '} P_t \left({\vec r}\ ' \right) 
W \left({\vec r}\ ' \to  {\vec r}  \right) 
 -  P_t \left({\vec r} \right) W_{out} \left( {\vec r} \right)
\label{masteru2d}
\end{eqnarray}
where the transition rates are given by the Metropolis choice
at temperature $T$ (the numerical data presented below correspond to $T=1$)
\begin{eqnarray}
W \left( \vec r \to \vec  r \ '  \right)
= \delta_{<\vec r, \vec r\ ' >} 
\  {\rm min} \left(1, e^{-  (U(\vec r \ ' )-U(\vec r ))/T } \right)
\label{metropolisu2d}
\end{eqnarray}
where the factor $\delta_{<\vec r, \vec r\ ' >}$
 means that the two positions
are neighbors on the two-dimensional lattice.
The random potential $U(\vec r)$ is self-affine with Hurst exponent $H=0.5$
\begin{eqnarray}
\overline{ \left[ U(\vec r) -U(\vec r \ ') \right]^2 }
\opsimeq_{ \vert \vec r - \vec r \ ' \vert \to \infty}
 \vert \vec r - \vec r \ ' \vert^{2H}
\label{correU2d}
\end{eqnarray}

On Fig. \ref{figu2d} (a), we show
 the corresponding probability distribution $Q_L(\Gamma_{eq})$
for various sizes  $10 \leq L \leq 80$ with a statistics 
of  $36.10^5 \geq n_s(L) \geq 4.10^4$  disordered samples. 

As shown by the log-log plots of Fig. \ref{figu2d} (b),
we find that the disorder-averaged value ${\overline \Gamma_{eq}(L)}$
 and the width $\Delta(L)$ of the distribution $Q_L(\Gamma_{eq})$
of the equilibrium barrier of Eq. \ref{defGammaeq}
involve the barrier exponent $\psi$
\begin{eqnarray}
 \overline{\Gamma_{eq}}(L) && \oppropto_{L \to \infty}  L^{\psi} \nonumber \\
\Delta(L)  &&  \oppropto_{L \to \infty} L^{\psi}
\label{psiboth}
\end{eqnarray}
of value
\begin{eqnarray}
\psi=H=0.5
\label{psiU2d}
\end{eqnarray}
These results are in agreement
 with scaling arguments on barriers \cite{Mar83,jpbreview}
and with the strong disorder renormalization approach of \cite{rgu2d}.

 \section{ Dynamics of the Sherrington-Kirkpatrick spin-glass model  }

\label{secSK}

\begin{figure}[htbp]
 \includegraphics[height=6cm]{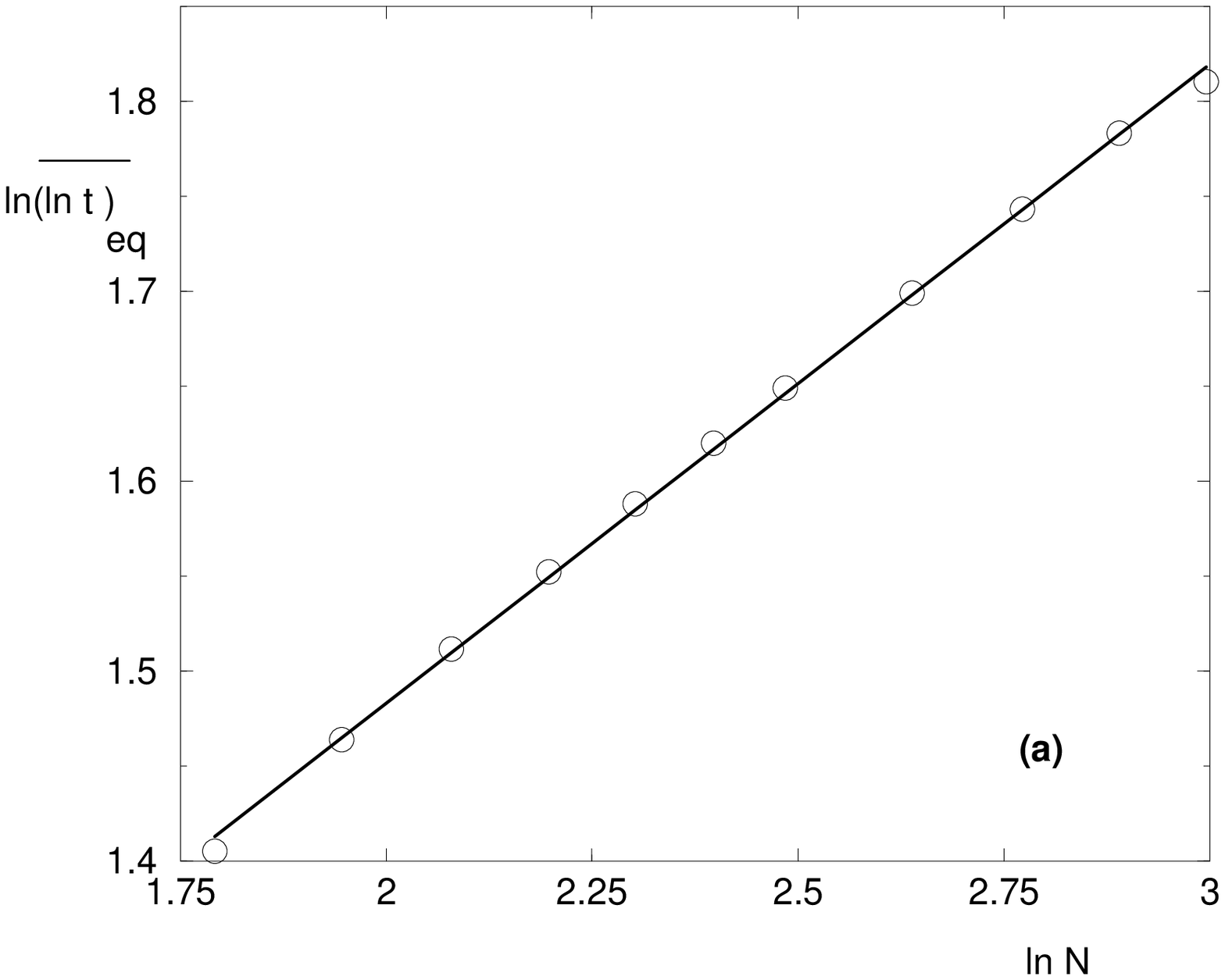}
\hspace{2cm}
 \includegraphics[height=6cm]{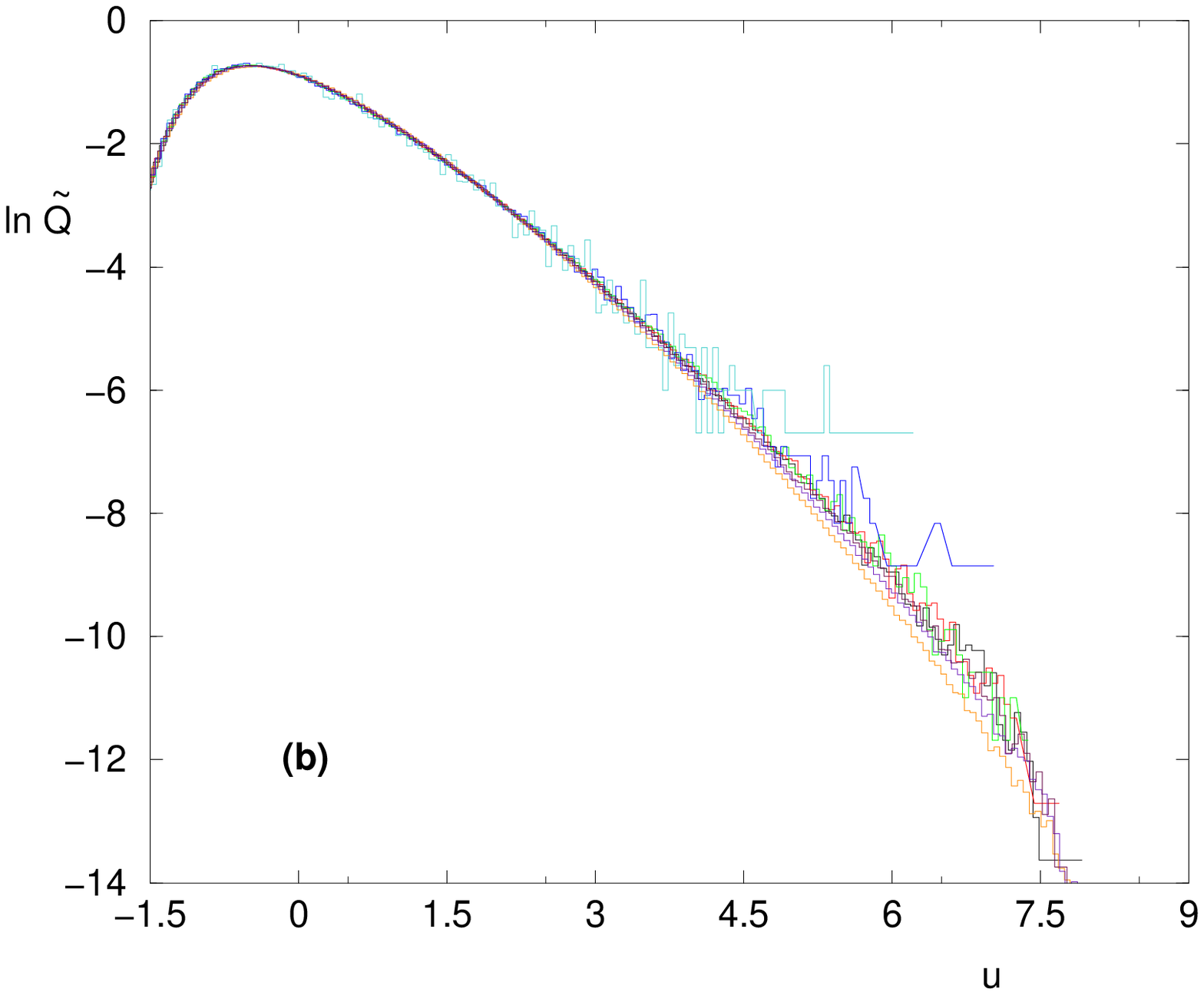}
\vspace{1cm}
\caption{ (Color on line) 
 Statistics of the equilibrium time $t_{eq}$ over the disordered samples
 for the Sherrington-Kirkpatrick spin-glass model of $N$ spins ($2^N$ configurations) :
(a) the log-log plot of the disorder-average $ \overline{\Gamma_{eq}}(L) $ as a function of $N$ for $6 \leq N \leq 20$ corresponds to the barrier exponent $\psi=1/3$ (Eq. \ref{psiSK})
(b) The rescaled probability distribution 
 ${\tilde Q}(u)$ of Eq. \ref{defQrescaled}, shown here for  $8 \leq N \leq 16$,
in log scale to see the tail of Eq. \ref{defeta} : 
the tail exponent is of order $\eta \simeq 1.36$.  }
\label{figSK}
\end{figure}

As an example of application to a many-body disordered system,
we consider in this section the of the 
Sherrington-Kirkpatrick spin-glass model where a configuration 
${\cal C}=\{S_i\}$ of $N$ spins $S_i =\pm 1$ has for energy \cite{SKmodel}
\begin{eqnarray}
 U = - \sum_{1 \leq i <j \leq N} J_{ij} S_i S_j
\label{defSK}
\end{eqnarray}
where the couplings are random quenched variables
of zero mean $\overline{J}=0$ and of variance $\overline{J^2}=1/(N-1) $.
The Metropolis dynamics corresponds to the master equation of Eq. \ref{master}
in configuration space with the transition rates
\begin{eqnarray}
W \left( {\cal C} \to  {\cal C} '  \right)
= \delta_{<{\cal C}, {\cal C}\ ' >} 
\  {\rm min} \left(1, e^{-  (U({\cal C} \ ' )-U({\cal C} ))/T } \right)
\label{metropolisSK}
\end{eqnarray}
where the factor $\delta_{<{\cal C}, {\cal C}\ ' >}$ 
means that the two configurations are related by a single spin flip.
The data presented below correspond to the temperature $T=0.5=T_c/2$.

In the conjugate gradient described in section \ref{secconjugate},
one can start from a random trial vector to begin the iterative method that will
 converge to the first excited eigenvector. 
However, in the case of spin models where $U$ is unchanged if one flips
 all the spins $S_i \to -S_i$, 
one knows that the largest relaxation time will correspond to a global flip of all
the spins. In terms of the quantum Hamiltonian associated to the dynamics
discussed in section \ref{secquantum}, this means that the ground state $\psi_0$
of Eq. \ref{psi0} is symmetric under a global flip of all the spins,
whereas the first excited state $\psi_1$ is anti-symmetric 
under a global flip of all the spins.
As a consequence, we have taken as initial trial eigenvector for the conjugate gradient method
the vector $\vert v >$ defined as follows : 
denoting ${\cal C}_{pref}=\{ S_i^{pref} \}$ and ${\widehat {\cal C}}_{pref}=\{ - S_i^{pref} \}$
the two opposite configurations where the ground state $\psi_0$ of Eq. \ref{psi0} is maximal,
one introduces the overlap between an arbitrary configuration ${\cal C}$ and ${\cal C}_{pref}$
\begin{eqnarray}
Q({\cal C},{\cal C}_{pref}) =  \sum_{i=1}^N S_i S_i^{pref}
\label{overlap}
\end{eqnarray}
and the vector
\begin{eqnarray}
v({\cal C}) = {\rm sgn } \left(Q({\cal C},{\cal C}_{pref}) \right) \psi_0( {\cal C})
\label{vinitialconjugate}
\end{eqnarray}
This vector is anti-symmetric under a global flip of all the spins and thus orthogonal
to the ground state $\psi_0$. Moreover, it has already a small Rayleigh quotient
(Eq. \ref{rayleigh}) because within each valley where the sign of the overlap is fixed,
it coincides up to a global sign with the ground state $\psi_0$ of zero energy.
So the non-zero value of the Rayleigh quotient of Eq. \ref{rayleigh} only comes from 
configurations of nearly zero overlap $Q$. As a consequence it is a good starting point
for the conjugate gradient method to converge rapidly towards
 the true first excited state $\psi_1$.

We have studied systems of $6 \leq N \leq 20$ spins (the space of configurations is of size $2^N$), with a statistics of $10^7 \geq n_s(N) \geq 1150$ of independent 
disordered samples to compute the probability distribution
$Q_N(\Gamma_{eq})$ of the largest barrier defined as
\begin{eqnarray}
\Gamma_{eq} \equiv \ln t_{eq}
\label{defbarrierflip}
\end{eqnarray}

As shown on Fig. \ref{figSK}(a), we find 
that the disorder averaged equilibrium barrier
scales as
\begin{eqnarray}
\overline{\Gamma_{eq}(N)} \equiv \overline{ \ln t_{eq}(N) } \oppropto_{N \to \infty} N^{\psi}
\ \ { \rm with } \ \ \psi \simeq 0.33
\label{psiSK}
\end{eqnarray}
This result is in agreement with theoretical predictions \cite{rodgers,horner}
and with most previous numerical measures \cite{young,vertechi,colbourne,billoire,janke}. It is also interesting to consider the sample-to-sample fluctuation exponent
$\psi_{width}$ that governs the width of the probability distribution of the
barrier
\begin{eqnarray}
\Delta(N) \equiv
 \left( \overline{\Gamma_{eq}^2}(N)
 - (\overline{\Gamma_{eq}}(N))^2\right)^{1/2} 
  \oppropto_{N \to \infty} N^{\psi_{width}}
\label{SKwidth}
\end{eqnarray}
Although the disorder-average value has been much studied numerically
\cite{young,vertechi,colbourne,billoire,janke}, the only measure 
of $\psi_{width}$ we are aware of, is given by Bittner and Janke \cite{janke}
\begin{eqnarray}
\psi_{width} \simeq 0.25
\label{widthjanke}
\end{eqnarray}
With our numerical data limited to small sizes $6 \leq N \leq 20$,
we see already the expected behavior of the disorder-average of Eq.
\ref{psiSK} as shown on Fig. \ref{figSK} (a), 
but we are unfortunately not able to measure
the exponent $\psi_{width}$ of Eq. \ref{SKwidth} from the variance.

However, as shown on Fig. \ref{figSK}(b), the probability distribution
$Q_N(\Gamma_{eq})$
convergences rapidly towards a fixed rescaled distribution ${\tilde Q}$
\begin{eqnarray}
Q_{N}(\Gamma_{eq}) \sim  
  \frac{1}{\Delta(N) } {\tilde Q} 
\left( u \equiv \frac{\Gamma_{eq} - \overline{\Gamma_{eq}}(N) }{\Delta(N) }
 \right)
\label{defQrescaled}
\end{eqnarray}
We find that the rescaled distribution ${\tilde Q} (u) $ presents at large argument
the exponential decay
\begin{eqnarray}
\ln {\tilde Q} (u) \oppropto_{u \to + \infty} - u^{\eta}
\label{defeta}
\end{eqnarray}
with a tail exponent of order
\begin{eqnarray}
\eta \simeq 1.36
\label{etaSK}
\end{eqnarray}
(on Fig. \ref{figSK}(b), a straight line would correspond to $\eta=1$.
Here we see a clear curvature indicating $\eta>1$. The value of Eq. \ref{etaSK}
has been estimated via a three-parameters fit 
$\ln {\tilde Q} (u) \simeq a- b u^{\eta}$ for the data in the range $u \geq 1$).
We are not aware of any theoretical prediction or any previous numerical measure
of this tail exponent $\eta$ to compare with.
However, it should have an interpretation in terms of rare events.
If the tail is due to rare samples that occur with some exponentially small
 probability of order $e^{- (cst) N^{\alpha}}$, but which present an anomalously large barrier of order $N^{\beta}$, the consistency equation for the powers of $N$ 
in the exponentials read,
using Eqs \ref{defQrescaled} and \ref{defeta}
\begin{eqnarray}
\left( \beta - \psi_{width} \right) \eta = \alpha
\label{rareSK}
\end{eqnarray}
We may now consider the contribution of various types of rare events :

(i) the anomalously ferromagnetic samples correspond to $\alpha=2$ (
with probability of order $e^{- (cst) N^2}$, the $N^2$ random variables 
$ {\tilde J}_{ij}$ will be all positive) and to $\beta=3/2$ 
(instead of being finite, the local field $h_i=\sum_j J_{ij} S_j $ 
on spin $S_i$ will be of order $N^{1/2}$).
If $\psi_{width}=1/3$, the corresponding tail exponent is $\eta= \frac{12}{7} =1.714$
which we have measured elsewhere \cite{firstpassagerg}
 for the case of the ferromagnetic Sherrington-Kirkpatrick model.
Since here we measure a significantly different value, we believe that the 
rare events dominating the tail for the spin-glass Sherrington-Kirkpatrick model
are not these ferromagnetic rare samples.

(ii) in a typical sample, the distribution $P(h)$ of the local field
extends down to $h=0$, with the linear behavior $p(h) \propto h$ as $h \to 0$)
\cite{anderson_houches,palmer,boettcher_KS}. However, with an exponentially
small probability of order $e^{- (cst) N}$, the $N$ local fields of the sample
will remain finite, i.e. bigger than some finite threshold $h_i \geq K$,
and the corresponding barrier will be anomalously large and of order $N$.
These rare samples, that have an 'anormalously strong spin-glass order',
in the sense that all local fields remain finite, thus correspond to the values $\alpha=1=\beta$ in Eq. \ref{rareSK}.
For instance, if $\psi_{width}=1/3$, 
the corresponding tail exponent reads $\eta= \frac{3}{2}$,
whereas if $\psi_{width}=1/4$, the corresponding tail exponent reads
 $\eta= \frac{4}{3}$.
Our measure of Eq. \ref{etaSK} corresponds to
\begin{eqnarray}
 \psi_{width} = 1 - \frac{1}{\eta} \simeq 0.26
\label{psifrometa}
\end{eqnarray}

A tentative conclusion would thus be the following :
at the small sizes that we can study, we cannot measure the width exponent
$ \psi_{width}$ from the variance, but we can measure the tail exponent $\eta$ that contains the information on $\psi_{width}$ if one can properly identify
 the rare events that dominate the tail.
In the spin-glass phase considered here, we believe that 
the rare events dominating the tail are the rare samples
described in (ii) that have an 'anormalously strong spin-glass order',
in the sense that all local fields remain finite, 
 so that our measure of the tail exponent of Eq. 
\ref{etaSK} would point towards the value of Eq. \ref{psifrometa}
for the width exponent, which is actually very close
 to the value of Eq. \ref{widthjanke}
measured by Bittner and Janke \cite{janke} from the variance for large sizes $N$.
These two indications suggest that $\psi_{width}$ could actually be strictly smaller
than the exponent $\psi=1/3$ governing the disorder-average value (Eq. \ref{psiSK}).
To the best of our knowledge,
 this question has never been raised for the barrier statistics,
but it has been much discussed for the statistics of the ground state energy
in the SK model (see \cite{palassini,bouchaud_KM,aspelmeier_BMM,aspelmeier,boettcher} and references therein), where the sample-to-sample exponent $\theta_{width}$ 
of the ground state energy (or the finite temperature free energy) is claimed
to be either $\theta_{width}=1/4$ or $\theta_{width}=1/6$, 
but is considered, in any case, to be
 smaller than the exponent $\theta_{av}=1/3$ 
that governs the correction to extensivity of the disorder average.
A natural question is also whether the values 
$\psi=1/3$ and $\psi_{width} \simeq 1/4$ found in the
 statistics of the dynamical barrier
are related to the exponents $\theta_{av}=1/3$ and $\theta_{width}$ 
that appear in the statistics of the ground state energy.

 \section{ Conclusion  }

In this paper, we have proposed to use the mapping between
 any master equation satisfying
detailed balance and a Schr\"odinger equation in configuration space
to compute the largest relaxation time $t_{eq}$ of the dynamics
via lowest non-vanishing eigenvalue $E_1=1/t_{eq}$ 
of the corresponding quantum 
Hamiltonian $H$ (the lowest eigenvalue being $E_0=0$).
This method allows to study the largest relaxation time $t_{eq}$   
{\it without simulating the dynamics }
by any eigenvalue method able
to compute the first excited energy $E_1$. 
In the present paper, we have used
the 'conjugate gradient' method (which is a simple iterative algorithm
related to the Lanczos method) to study the statistics of the equilibrium time
in two disordered systems :

(i) for the random walk in a two-dimensional self-affine potential of Hurst exponent $H$ 

(ii) for the dynamics of the Sherrington-Kirkpatrick spin-glass model of $N$ spins.

The size of vectors used in the 'conjugate gradient' method is the size
${\cal N}_C$ of the configuration space for the dynamics: for instance
it is ${\cal N}_C=L^2$ for the case (i) 
of a single particle on the two-dimensional square $L \times L$ and 
it is ${\cal N}_C=2^N$ for (ii) containing $N$ classical spins. 
We have shown here that the conjugate gradient method was sufficient 
to measure the barrier exponents for these two models, but it is clear
that it will not be sufficient for spin models in dimension $d=2$ or $d=3$
where the size of the configuration space grows as 
$2^{L^d}$, and that it should
be replaced by a quantum Monte-Carlo method to evaluate $E_1$.
For instance for the dynamics of the pure two dimensional
 Ising model at criticality
studied in \cite{Night_Blo}, 
the conjugate-gradient method used for squares $L^2$
of sizes $L \leq 5$ has been replaced for bigger sizes $5 \leq L \leq 15$
by a quantum Monte-Carlo method appropriate to compute
excited states  \cite{bernu}.
We thus hope that the same strategy will be useful
 in the future to compute the equilibrium time
of disordered spin models in dimension $d=2$.

\section*{Acknowledgements }

It is a pleasure to thank A. Billoire, J.P. Bouchaud, A. Bray and M. Moore
for discussion or correspondence on the statistics
 of dynamical barriers in mean-field spin-glasses.

\label{secconclusion}

\end{document}